\documentclass[aps,a4paper,10pt,twocolumn,nofootinbib]{revtex4} % eqsecnum
\usepackage{datetime}
\usepackage{amsmath}
\usepackage{amsfonts}
\usepackage{amsfonts}
\usepackage{mathrsfs}
\usepackage[mathscr]{euscript}
\usepackage[dvips]{graphicx}
\usepackage{fancyhdr}
\usepackage{colordvi}
\usepackage[hypertex=true]{hyperref}
\usepackage{epsfig}
\usepackage{color}
\usepackage{bm}

\def\overstrike#1#2{{\setbox0\hbox{$#2$}\hbox to \wd0{\hss
    $#1$\hss}\kern-\wd0\box0}}

\newcommand{\XDOI}[1]{\href{http://dx.doi.org/#1}{doi:#1}}

\newdateformat{yymmdddate}{\THEYEAR/\twodigit{\THEMONTH}/\twodigit{\THEDAY}}

\begin{document}
\title{Maxwell's Fishpond}
\author{Paul Kinsler}
\email{Dr.Paul.Kinsler@physics.org}
\author{Jiajun Tan}
\author{Timothy  C. Y. Thio}
\author{Claire Trant}
\author{Navin Kandapper}
\affiliation{
  %Blackett Laboratory, Imperial College,
  Department of Physics, 
  Imperial College London,
  Prince Consort Road,
  London SW7 2AZ, 
  United Kingdom.
}

\lhead{\includegraphics[height=5mm,angle=0]{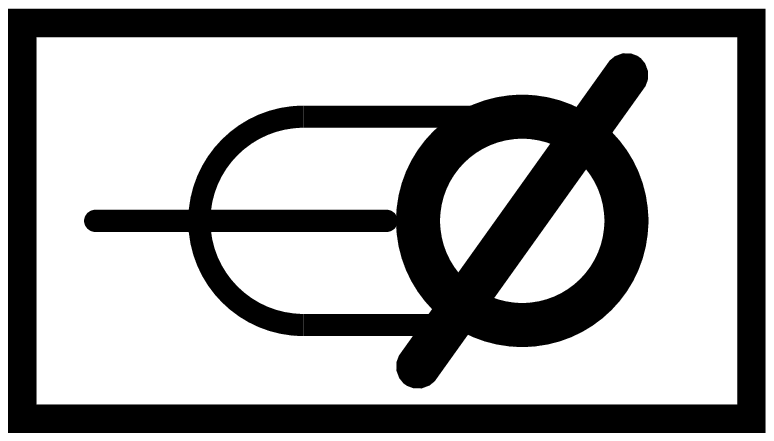}~~FISHPOND}
\chead{Maxwell's Fishpond}
\rhead{
\href{mailto:Dr.Paul.Kinsler@physics.org}{Dr.Paul.Kinsler@physics.org}\\
\href{http://www.kinsler.org/physics/}{http://www.kinsler.org/physics/}
}
%\lfoot{\thesection . \thesubsection; ~~~~ (\yymmdddate\today:\currenttime) 
% \href{http://localhost/}{(0)}}
%\rfoot{{\large \emph{Not for redistribution}}}

\begin{abstract}

Most of us will have at some time thrown a pebble into water, 
 and watched the ripples spread outwards and fade away.
But now there is also a way to reverse the process, 
 and make those ripples turn around and reconverge again,
 ...
 and again, 
 and again.
To do this we have designed the Maxwell's Fishpond, 
 a water wave or ``Transformation Aquatics''
 version of the 
 Maxwell's fisheye lens \cite{Tyc-HSB-2011njp,Luneberg-MTO}.
% that is now well-known from transformation optics.
These are transformation devices where wave propagation
 on the surface of a sphere is modelled
 using a flat device with spatially varying properties.
And just as for rays from a point source on a sphere, 
 a wave disturbance in a Maxwell's fisheye or Fishpond spreads out at first,
 but then reforms itself at its opposite (or complementary) point.
Here we show how such a device can be made for water waves, 
 partly in friendly competition
 with comparable electromagnetic devices \cite{Ma-SOTL-2011njp}
 and partly as an accessible and fun demonstration
 of the power of transformation mechanics.
To the eye, 
 our Maxwell's Fishpond was capable of reforming a disturbance
 up to five times, 
 although such a feat required taking considerable care, 
 close observation, 
 and a little luck.\\
~\\
What can \emph{you} see in the video at 
 \href{http://www.qols.ph.ic.ac.uk/\~kinsle/files/MFishpond/}{http://www.qols.ph.ic.ac.uk/$\tilde{~}$kinsle/files/MFishpond/} ?

\end{abstract}

\pacs{47.35.Bb,43.20.Dk,43.20.El,01.50.My}

% 43.20.Dk Ray acoustics 
% 43.20.El  Reflection, refraction, diffraction of acoustic waves (see also 43.30.Es)
% 47.35.Bb  Gravity waves (pknb - under hydrodynamic waves)

% 43.30.Es Velocity, attenuation, refraction, and diffraction in water, Doppler effect (pknb - underwater)
% 01.50.My  Demonstration experiments and apparatus

% 42.30.Va Image forming and processing  
% 07.60.-j Optical instruments and equipment

% Keywords: maxwell, fisheye, fishpond, water waves, transformation, imaging

\date{\today}
\maketitle
\thispagestyle{fancy}

\noindent{\emph{This is an updated preprint
 of the published article\footnote{\textbf{Statement required by the publisher of the EJP:}
 This is an author-created, un-copyedited version of an article
  accepted for publication in the European Journal of Physics. 
 IOP Publishing Ltd is not responsible for any errors or omissions
  in this version of the manuscript or any version derived from it. 
 The definitive publisher-authenticated version is available online
 at doi:10.1088/0143-0807/33/6/1737.}
}  \\
Kinsler et al., Eur. J. Phys. \textbf{33}, 1737 (2012).\\
\small{\url{http://iopscience.iop.org/0143-0807/33/6/1737}}}

%
%\tableofcontents

%
% =======================================================================
\section{Introduction}\label{S-intro}

Transformation Optics \cite{Kildishev-S-2011pu} 
 and
 Transformation Acoustics \cite{Chen-C-2010jpd}
 are powerful new techniques used to design 
 \emph{transformation devices}, 
 which usually tend to the exotic -- 
 invisibility or event cloaks \cite{McCall-FKB-2011jo}, 
 illusion generators, 
 and so on.
Unfortunately, 
 none are useful as demonstration devices accessible
 to non-specialists.

Here, 
 in contrast to this typical situation, 
 we describe how to make a transformation device (T-device)
 that fits on a tabletop and which controls water waves
 visible to the naked eye.
Although T-devices such as cloaks cannot be made with simple isotropic
 materials, 
 one important type can: 
 one whose design principle transforms
 from waves travelling on the surface of a uniform sphere, 
 to waves travelling on a flat disk.
The transformation used preserves the properties 
 of the original wave propagation -- 
 that of circular trajectories of equal circumference --
 by making the disk properties non-uniform:
 here, 
 we make a shallow pond with varying depth.

In optics, 
 such a device is known as the 
 Maxwell's fisheye lens \cite{Tyc-HSB-2011njp,Luneberg-MTO}.
It has a long history, 
 originally being proposed in a problem set
 in the 1850's \cite{Maxwell-1853-fisheye,Maxwell-1854-fisheye-soln}.
Until recently an obscure theoretical curiosity, 
 the device was brought to wider attention by 
 controversial claims that 
 such a device could generate perfect optical images \cite{Leonhardt-2009njp}.
Leaving aside that debate
 (see \cite{Blaikie-2011njp} for a recent critical summary), 
 Leonhardt usefully noted that actually building the device 
 becomes possible if you take 
 only the central portion and surround it by a mirror 
 \cite{Leonhardt-2009njp}.
This trick also makes the water wave version easier to build.
But what makes a Maxwell's fisheye lens --
 or its Fishpond counterpart -- 
 interesting?

As in an ordinary pond, 
 a pointlike wave source anywhere on the surface
 of a Maxwell's Fishpond
 generates an outgoing set of ripples.
However, 
 in a Maxwell's Fishpond, 
 the ripples do not just spread out and disperse, 
 they \emph{also} converge on the opposite side,
 before again diverging
 and travelling back to the start where they again reconverge, 
 and repeat this process until they eventually dissipate.
This is just as light does
 in the optical Maxwell's fisheye lens \cite{Kinsler-F-2010njp-fisheye}, 
 and is exactly what rays or waves confined 
 on the surface of a sphere would do.
More generally, 
 the Maxwell's fisheye 
 is one of a more general class of classical \emph{transformation optics}
 devices \cite{Tyc-HSB-2011njp}; 
 and these others, 
 such as the Eaton or Luneberg lenses, 
 will also have accessible transformation aquatics equivalents
 based on water waves --
 just as they do in the more technologically exotic 
 field of plasmonics \cite{Kadic-DGE-2011jmo,Zentgraf-LMVZ-2011nn}.

Here we will derive the water depth profile
 needed to make a Maxwell's Fishpond in the shallow water limit, 
 and compare that profile to a simple approximation
 using a shallow spherical dome.
Although the approximation can work surprisingly well, 
 our accurate device does better, 
 and can hint at -- 
 at least to the eye -- up to \emph{five} successive refocussings!
More rigorously, 
 we also present simulation results indicating how the device 
 works in practise, 
 as well as two experimental schemes set up with relatively
 little demands on equipment.

Contributions to this work were as follows:
PK conceived the Maxwell's Fishpond idea, 
 and built the first crude prototype.
He also designed the version used here, 
 but with students NK and TT shadowing that design process.
NK and TT did the first experiments, 
 and CT and JT followed next; 
 all four writing reports and giving presentations as part of their coursework.
PK was the primary author of this paper, 
 assisted by material from the student reports, 
 he also did all the computer simulation work.
CT, JT, and TT also assisted in the preparation
 of this final manuscript.

%
% =======================================================================
\section{Fisheye, Fishpond}\label{S-fish}

The Maxwell's fisheye concept is based on mimicking 
 the properties of ray trajectories on the surface of a sphere
 using a flat surface with spatially modulated properties.
This is interesting, 
 because on a sphere any set of rays emitted from a point
 follow their individual ``great circle'' geodesics, 
 and so will automatically converge on the 
 exact opposite side of the sphere.
Thus, 
 any flat T-device version should also have this property -- 
 rays diverging from \emph{any} point would automatically 
 focus at the complementary point of the plane.
Thus, 
 both on the sphere and in the fisheye, 
 an object at any point is guaranteed to form an image; 
 this is most certainly not the case in ordinary imaging systems.
Further, 
 the rays would then re-diverge before converging again; 
 in an ideal ray device, 
 these image reformations would continue forever.

\begin{figure}
\centering
\includegraphics[angle=-0,width=0.80\columnwidth]{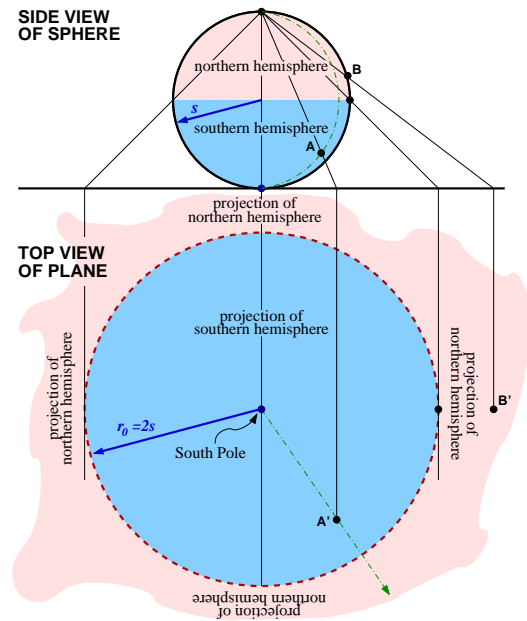}
\caption{
The sphere-to-plane fisheye projection can be imagined by considering
 a transparent sphere with a light source placed at the north pole $N$, 
 objects on the sphere then cast a shadow on the plane matching 
 the projection.
The southern hemisphere of the sphere of radius $s$ 
 becomes a finite disk of radius $r_0=2s$, 
 whereas the northern hemisphere becomes the entire plane 
 that remains \emph{outside} the disk.
Lines from pole-to-pole 
 (meridians, or lines of longitude)
 on the sphere become radial lines
 (see e.g. the dot-dashed line), 
 circles on the sphere parallel to the equator 
 (parallels, or circles of latitude) 
 become projected circles whose size depends on how far north or south 
 of the equator (dashed line) they are.
}
\label{fig-projection2D}
\end{figure}

To achieve the transformation from a spherical device 
 with its curved surface,
 to a flat one, 
 we use a stereographic projection.
%an \emph{azimuthal} conformal projection.
Imagine a sphere sitting with its south pole on a flat sheet, 
 as shown on the upper part of fig. \ref{fig-projection2D}.
Then any point (e.g. $A$ or $B$) on the sphere is mapped onto the sheet
 by following a straight line from the north pole, 
 through $A$ (or $B$), 
 and onward until it intersects the sheet at $A'$ (or $B'$).
In this way the curved southern hemisphere maps onto a disk on the flat sheet
 centered on the south pole.
The  northern hemisphere is mapped to points further away; 
 with points very near the north pole being extremely remote, 
 and the north pole itself having to be omitted.

%
% http://www.progonos.com/furuti/MapProj/Normal/TOC/cartTOC.html
%

Of course, 
 although we would like to make a fisheye (or Fishpond)
 based on this projection,
 we do not want one that is infinitely big.
We therefore follow Leonhardt \cite{Leonhardt-2009njp}
 and place a mirror at the equator, 
 confining all ray paths to the southern hemisphere, 
 and so confining all projected rays inside a circle
 with twice the sphere's radius.
Since both hemispheres have the same properties, 
 the ray properties are preserved -- although the 
 great circles are now folded back on themselves
 and have a kink where they are reflected, 
 they still are guaranteed to form a image of any point.

The process of opening out and flattening the surface
 of a sphere into an equivalent sheet, 
 as if it were a map projection used for an atlas, 
 has an important feature.
Regions near the equator are stretched and expanded, % greatly,
 while those near the south pole are only slightly changed.
Mathematically we can define 
 a complex quantity $z = x + \imath y$, 
 where $x, y$ represent the Cartesian coordinate on the plane.
This means that any point $(X,Y,Z)$ 
 (or at angles $\theta, \phi$)
 on the unit sphere
 is projected (or ``mapped'') down onto
~
\begin{align}
  z 
&= 
  \frac{X+\imath Y}{1-Z} 
&=
  \exp \left( \imath \phi \right)
  \cot\left( \theta/2 \right)
.
\end{align}
This means that a given line element $dS^2 = dX^2 + dY^2 + dZ^2$ on the sphere
 is transformed into a line element in the plane $dR^2=dx^2 + dy^2$
 that progressively lengthens as we move towards the equatorial perimeter 
 at $r=1$.
~
\begin{align}
  \frac{2}{1 + r^2}
  dR^2    %\left( dx^2 + dy^2\right)
&=
  dS^2 
.
\end{align}
 where $r^2 = x^2 + y^2$, 
 and we can also note that angles on the sphere 
 are preserved when projected onto the plane.

This length transformation 
 means that an object (or ray)
 travelling at a fixed speed on the sphere
 will have a projection on the plane that travels faster
 the closer it gets to the north pole.
Thus a fixed object speed $v_0$ on the sphere
 is projected onto the disk as a radially varying 
 velocity profile $v(r)$.
Because of the way the projection works --
 or the line element $dS$ converts to $dR$ --
 the velocity profile $v(r)$
 for a sphere of radius $r_0/2$
 is projected onto its counterpart disk of radius $r_0$
 as 
~
\begin{align}
  v(r)
&=
  v_0 \left[1 + \left(\frac{r}{r_0}\right)^2 \right]
.
\label{eqn-fisheye-v}
\end{align}
Any disk which transports objects or waves with this velocity profile
 will be a T-device representing a sphere.
Note that 
 this velocity profile has a counterpart in a 
 gradient-index version of Snell's law
 that steers propagating rays 
 so that they match the paths that follow from 
 projections of the great circle paths on the sphere. 
Indeed, 
 from a mathematical perspective we might have expected this, 
 because the projection preserves angles, 
 which also means that the device can be made with isotropic materials.

% (see e.g. \cite{Righini-RST-1972ao})

%
% -----------------------------------------------------------------------
\subsection{Optics}\label{S-fish-optics}

In optics, 
 one can actually make a thin spherical shell that
 will guide light inside it
 (see e.g. \cite{Righini-RST-1972ao}), 
 just like the spherical reference device for a planar fisheye lens.
But 
 to obtain a design for the flat Maxwell's fisheye lens
 we need to design an optical device 
 that has the light speed profile defined in eqn. \eqref{eqn-fisheye-v}
 by modulating the refractive index.
If starting with a shell with refractive index $n_0$
 (and hence speed of light $c' = c/n_0$),
 we can convert this to a disk with a radially varying 
 refractive index profile
~
\begin{align}
  n(r)
&=
  \frac{n_0}
       {1 + \left(r/r_0\right)^2}
,
\label{eqn-fisheye-n}
\end{align}
where $n_0$ is the maximum refractive index we can achieve, 
 and $r_0$ is our desired radius scale.
In the original fisheye lens, 
 $r$ was unbounded, 
 causing the device to need unrealistically small values of $n$
 at large radii $r$.
The introduction of an equatorial mirror, 
 as discussed above, 
 circumvents this restriction; 
 and for $n_0 \ge 2$ the minimum refractive index required
 is always $\ge 1$.
Electromagnetic Maxwell's fisheye lenses have been made
 (e.g. \cite{Ma-SOTL-2011njp,Smolyaninova-SKS-2010ol,Gabrielli-LL-2010arxiv,Gabrielli-L-2011njp}), 
 but require significant technological skill
 to build and investigate.
Hence our interest in water waves, 
 which gives a much wider audience
 access to these interesting devices.

%
% -----------------------------------------------------------------------
\subsection{Water waves}\label{S-fish-pond}

We want to make a Fishpond, 
 not a fisheye; 
 water waves are easily visible, 
 intuitive, 
 low-tech,
 and are accessible and safe for a wide variety of ordinary people.
Nevertheless, 
 experimental water wave systems can still be used
 as models for 
 quite a surprising variety of phenomena: 
 e.g. 
 event horizons
 and Hawking radiation \cite{Jannes-PCMMR-2011jpc,Rousseaux-MMPL-2008njp}
 and
 neutron star collapse \cite{Foglizzo-MGD-2012prl}.

To obtain a design for such a Maxwell's Fishpond
 we need only work out how to design a device 
 that has the speed profile defined in eqn. \eqref{eqn-fisheye-v}.
In general, 
 water waves can have a complicated and nonlinear behaviour, 
 so constructing a general fishpond will be either very difficult 
 or impossible.
But there is an important subset of water waves
 for which we \emph{can} get a simple solution -- 
 those waves that occur in very shallow water.

For water of a constant depth 
 that is significantly less than a wavelength, 
 the wave speed for small waves is simply \cite{Mayo-1997tpt}
~
\begin{align}
  v_w
&=
  \sqrt{gd}
\label{eqn-fishpond-water-c}
,
\end{align}
 where $g=9.81$m/s$^2$ is the gravitational acceleration
 and d is the water depth.
In this extreme limit, 
 other factors such as the wave amplitude and wavelength
 no longer matter, 
 and we can control \emph{any} suitable wave with the same
 depth modulation.
Further, 
 as long as $d(r)$ varies slowly over wavelength scales, 
 we can use the formula 
 to describe waves travelling across a varying depth profile $d(r)$.

If the centre of the fishpond has depth $d_0$,
 the water wave speed there is $v_w(0) = \sqrt{g d_0}$.
Thus the radial wave velocity profile will be 
~
\begin{align}
  v_w(r)
&=
  v_w(0)
  \sqrt{\bar{d}}
,
\label{eqn-fishpond-v}
\end{align}
 where $\bar{d}(r) = d(r)/d_0$ is the relative depth profile.

\begin{figure}
\includegraphics[angle=-0,width=0.85\columnwidth]{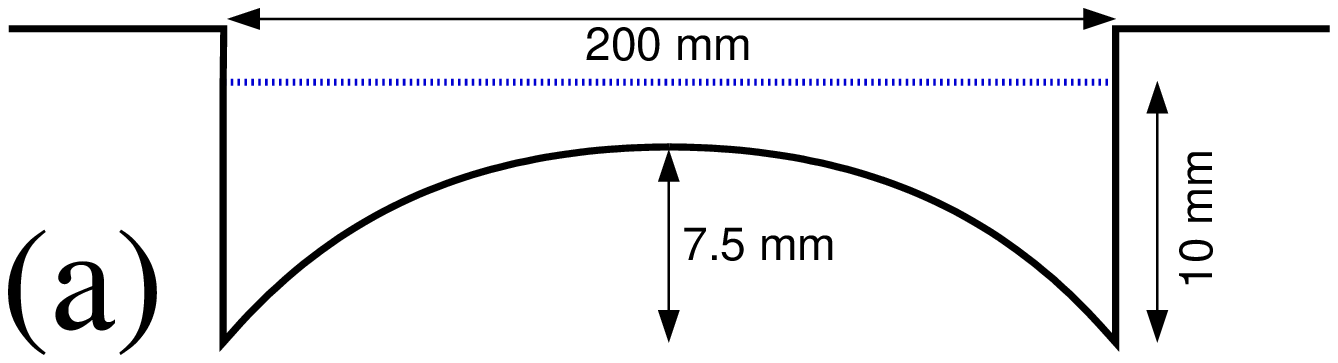}
\includegraphics[angle=-0,width=0.85\columnwidth]{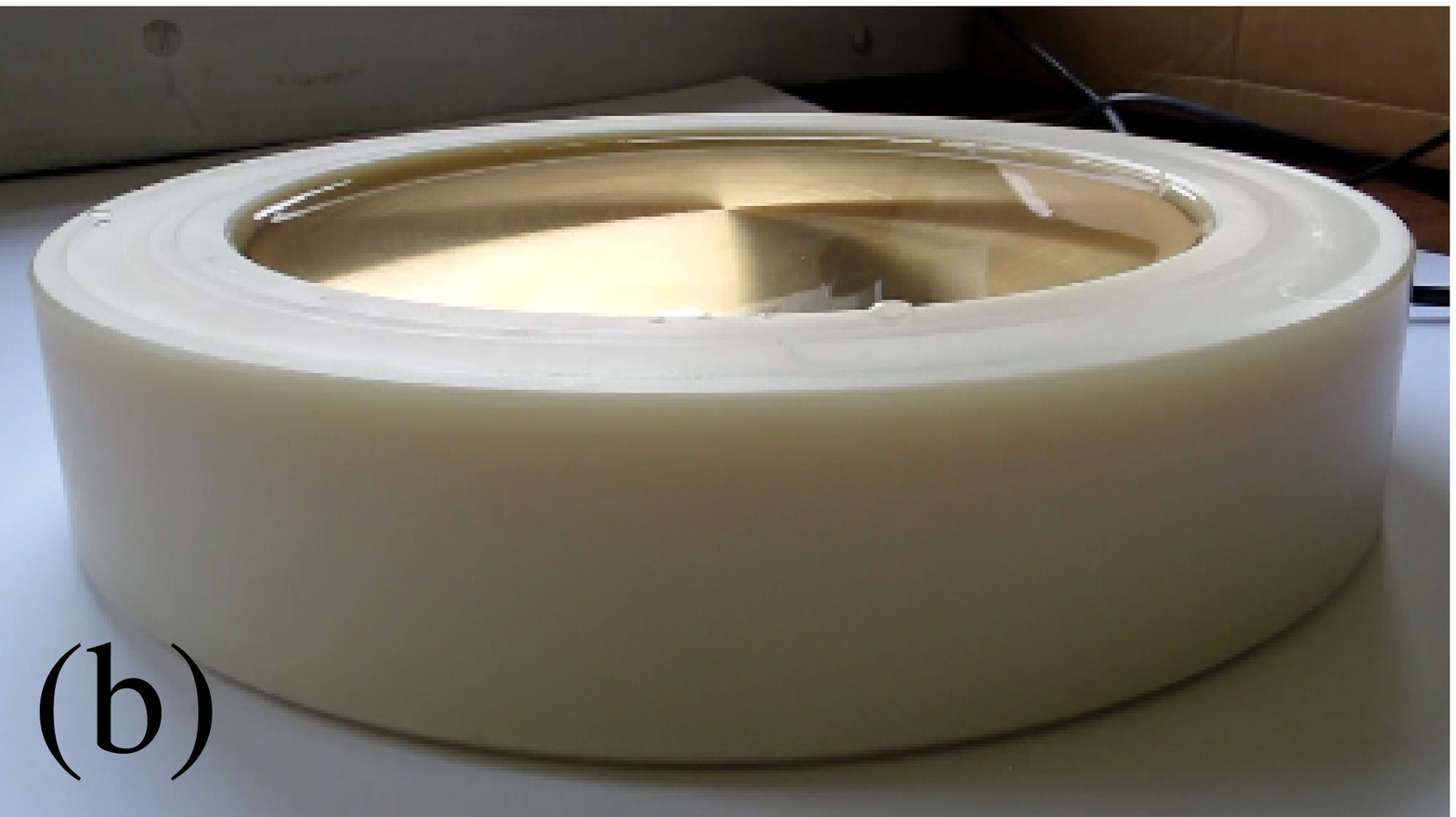}
\caption{
The Maxwell's Fishpond.
(a) A cross section
 with the vertical scale grossly exaggerated
 for clarity.
The indicated dimensions are those used for our actual device, 
 but heights or widths can be rescaled freely -- 
 subject to the proviso that ripples will have a wavelength
 longer than the maximum depth.
In fact, 
 an even shallower Fishpond would provide a better match to this criterion,
 but since water has a significant surface tension, 
 this makes covering the centre region problematic.
(b) A photograph of our device.
}
\label{fig-fishpond-made}
\end{figure}

Thus, 
 comparing eqns. \eqref{eqn-fisheye-v} and \eqref{eqn-fishpond-v}
 we see that to match the two velocity profiles
 we need to have
~
\begin{align}
  \sqrt{\bar{d}(r)}
&=
  \left[1 + \left(r/r_0\right)^2 \right]
\\
  \bar{d}(r)
&=
  \left[1 + \left(r/r_0\right)^2 \right]^2
\\
&=
  1 + 2 \left(r/r_0\right)^2 + \left(r/r_0\right)^4
\label{eqn-fishpond-vcf}
.
\end{align}
The parameters we chose for our Fishpond
 were based on an assumed water wavelength of about 20mm;
 the result can be seen in fig. \ref{fig-fishpond-made}.

% fishpond-diagram-3
% fishpond-photo

One might also construct other types of geodesic lens
 \cite{Luneberg-MTO,Sarbort-T-2012jo,Tyc-HSB-2011njp}
 using water waves,
 or other types of acoustic waves \cite{Bramhavar-PNENM-2011prb},
 as discussed in the appendix at \ref{S-fish-EatonEtc}
 and \ref{S-fish-other}.

%
% =======================================================================
\section{Modelling}\label{S-model}

We tested our design using computer simulations for a variety of cases
 ranging from those applicable to the ideal Maxwell's Fishpond
 and approximate Fishponds, 
 to a full finite element simulation for our Fishpond device.
For the idealized comparisons, 
 we used the fact that the fisheye and Fishpond
 behave in an essentially identical manner, 
 once the distinctions between the polarizable EM field
 and scalar water waves have been accounted for.
This means that since an ideal Fishpond has the same properties 
 as an ideal fisheye lens, 
 FDTD simulations of Maxwell's equations for the fisheye lens
 will indicate the behaviour of an ideal Fishpond.

\begin{figure}
\includegraphics[angle=-0,width=0.42\columnwidth]{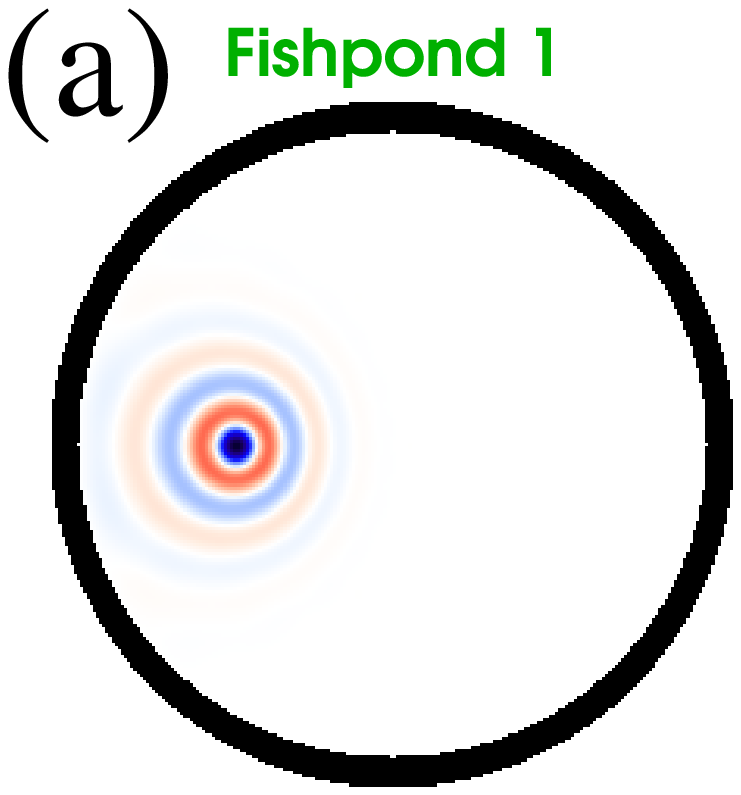}
\includegraphics[angle=-0,width=0.42\columnwidth]{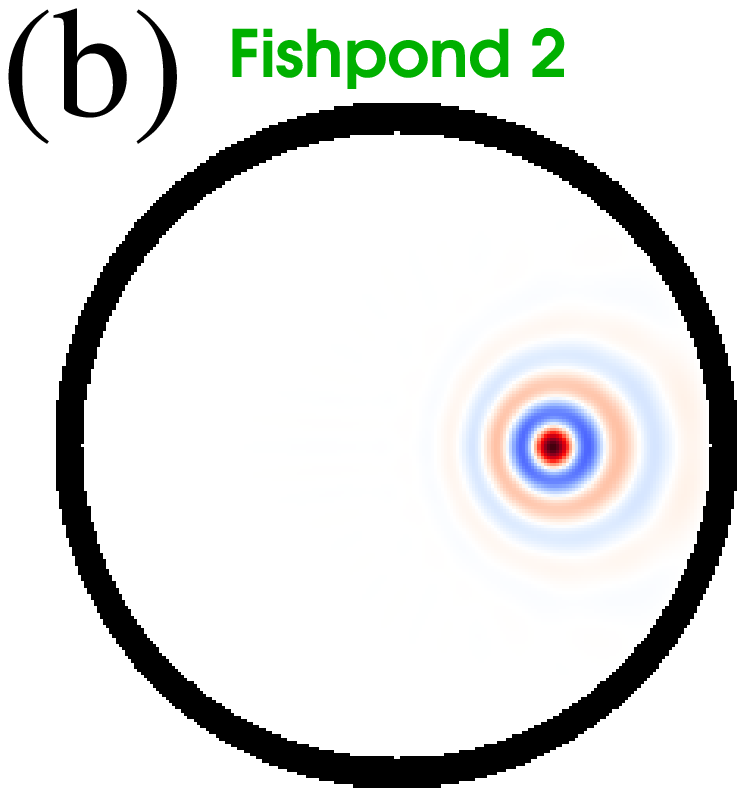}
\includegraphics[angle=-0,width=0.42\columnwidth]{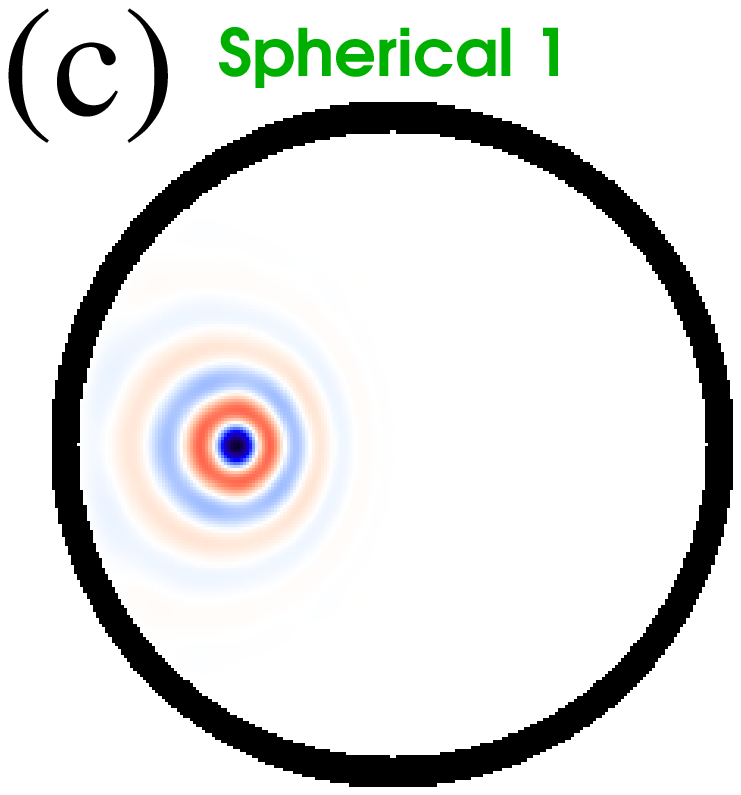}
\includegraphics[angle=-0,width=0.42\columnwidth]{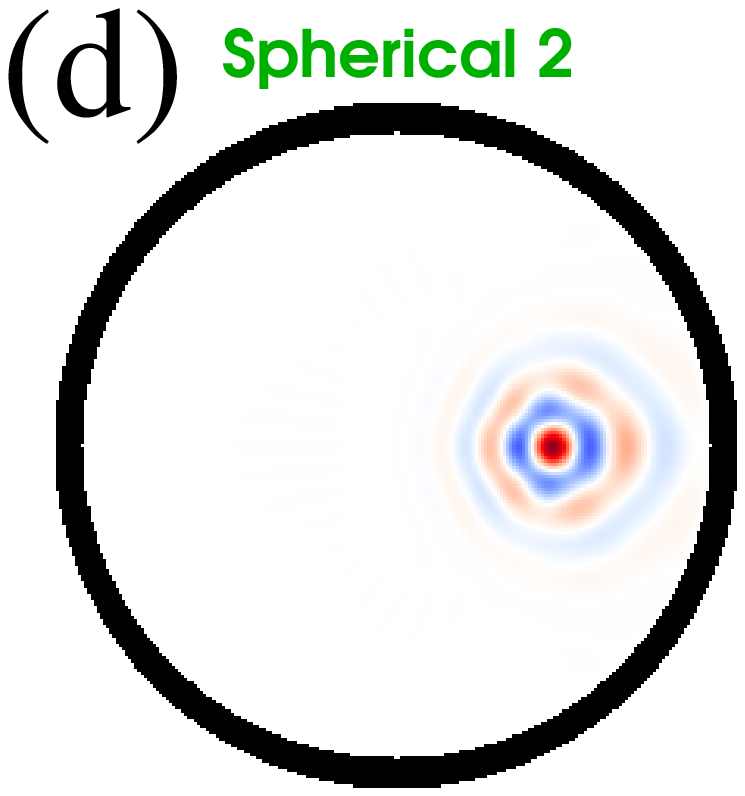}
\includegraphics[angle=-0,width=0.42\columnwidth]{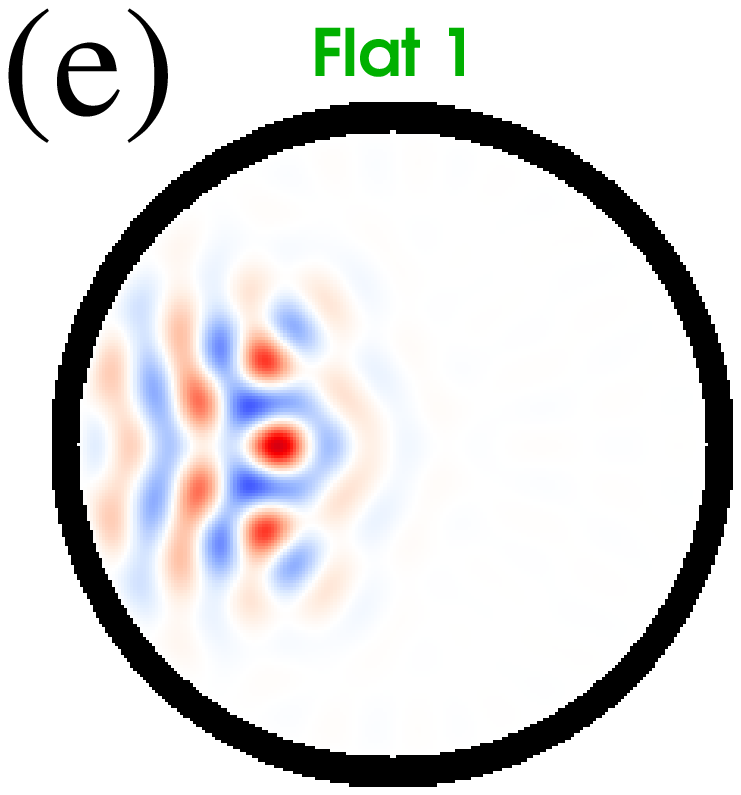}
\includegraphics[angle=-0,width=0.42\columnwidth]{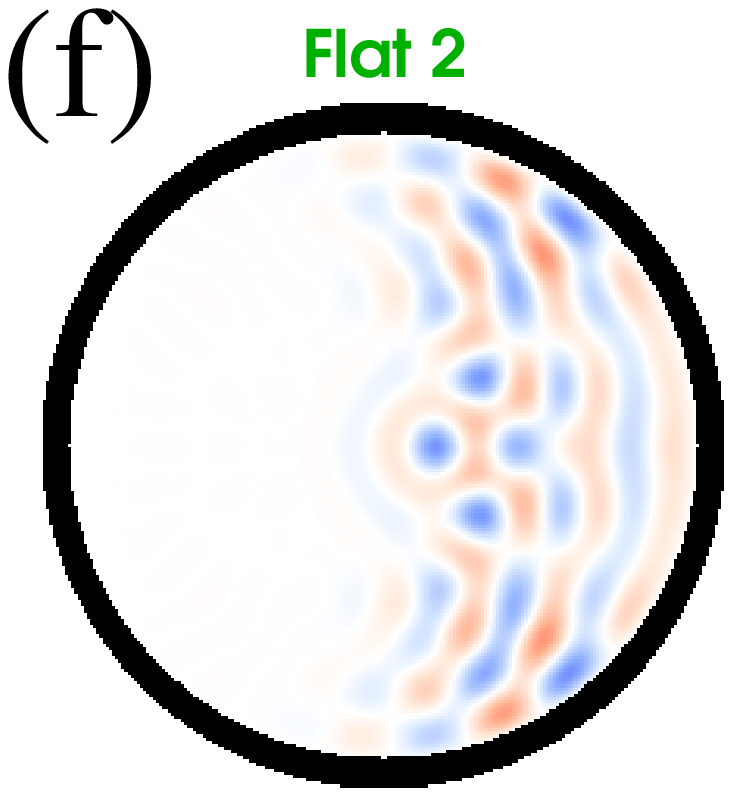}
\caption{
Snapshots from simulations representing an ideal Maxwell's Fishpond (a,b), 
 an approximate Fishpond with a spherical-cap (SC) depth profile (c,d), 
 and an ordinary flat-bottomed fishpond (e,f).
The upper frames (a, b) show the fisheye/pond wave patterns
 near the first and second refocussing times; 
 the middle ones (c,d) the approximate SC pond,
 and the lower two (e,f) show the non-focussing behaviour 
 of a flat-bottomed pond when the ripples are not started
 at the exact centre.
To the eye, 
 the second Fishpond reformation (b) 
 is essentially identical to the original source wave.
}
\label{fig-fishpond-snapshots}
\end{figure}

To get an initial estimate of the importance of the correct depth profile, 
 we used MEEP \cite{Oskooi-RIBJJ-2010cpc} 
 FDTD simulations of Maxwell's equations.
This approach was taken because we already had such EM simulations running, 
 and because the MEEP software 
 is flexible, open source, and freely available.
All that is required is to 
 converted our chosen depth profiles
 back into a a refractive index profile
 using the reverse of the process that led to eqn. \eqref{eqn-fishpond-vcf}.
Sample MEEP control files are available in appendix \ref{S-meepctl}.
As well as the exact Maxwell's Fishpond, 
 and amongst other variations, 
 we modelled an approximate depth profile
 based on a shallow spherical cap (SC).
It turns out that as long as the correct 1:4 ratio
 of minimum to maximum depths is maintained, 
 this worked remarkably well.

In fig. \ref{fig-fishpond-snapshots} we can see the simulation results 
 equivalent to our shallow water wave model, 
 showing snapshot pairs that demonstrate the image reformation properties.
We see that the ideal Maxwell's fisheye lens and Fishpond
 will give accurate refocussing
 (fig. \ref{fig-fishpond-snapshots}(a,b)), 
 and this is repeated very many times before the performance starts to degrade
 due to the dispersion caused by how different wavelengths
 interact with the finite-sized geometry.
 (see e.g. \cite{Kinsler-F-2010njp-fisheye}).
Next, 
 the simulations matching the approximate 
 domed pond profile, 
 do quite well 
 (see fig. \ref{fig-fishpond-snapshots}(c,d)), 
 but with some distortion clearly evident on the second reformation.
However,
 the chosen ``best reformation'' snapshots of the domed pond flatter slightly,
 as the frames before and after shown a significant ellipticity; 
 and the third reformation (not shown),
 whilst still giving a localised wave bunch, 
 has lost its concentric-ring character.
Finally, 
 simulations of a flat-bottomed pond
 (see  fig. \ref{fig-fishpond-snapshots}(e,f))
 show only a poor attempt at a first focus, 
 followed by a rapid evolution towards an apparently random pattern
 with no reformations apparent at all.
Other simulations including a variety of strengths of non-radial distortion
 of the Fishpond depth profile
 were also performed.
Depth variations of about 10\% away from the exact profile
 do give tolerable results
 for the first few image reformations,
 but the distortion % induces an unwanted ellipticity; 
 strongly degrades the beautiful concentric-ring character
 of the exact Maxwell's Fishpond.

Unfortunately, 
 it is hard to build a real water wave device that works
 perfectly
 in the shallow water limit.
Most notably, 
 we will expect to see some residual dispersion, 
 due to the depth-dependent speeds of different wavelength ripples
 and the effects of surface tension; 
 this is discussed later in section \ref{S-discuss}.
Thus the brief reformations expected in the ideal case 
 will blur out into longer process
 as different wave components refocus at different times, 
 and for the shorter $\lambda$ waves, 
 the depth profile will be less perfect.

More realistic finite element simulations
 were also done
 using the open source simulator OpenFOAM \cite{OpenFOAM}
 with the {interMixingFoam} engine
 on a fast desktop PC.
Despite computational constraints, 
 the simulations gave good results, 
 with the effect of dispersion demonstrated, 
 as can be seen in the appendix at \ref{S-model-OpenFOAM}.

%
% =======================================================================
\section{Experiments}\label{S-expt}

To make the Maxwell's Fishpond, 
 our departmental Mechanical Workshop
 machined us a brass insert with the 
 necessary depth profile (see fig. \ref{fig-fishpond-made})
 and mounted this in a nylon ring, 
 with a small notch to indicate the preferred water level.
The choice of brass and nylon was based on convenience, 
 not necessity, 
 any waterproof materials could suffice.
Our brass insert was not perfectly smooth, 
 but was machined to tolerances of much less than 1mm.

%-----------------------------------------------------------------------
%\subsection{Equipment setup} 

Two schemes for obtaining quantitative data were used:
 one by NK and TT, 
 the other by CT and JT.
Both were influenced by the fact that 
 although viewing the Fishpond directly 
 gives a very strong impression of how well it works, 
 this human perception does not translate easily
 into objective experimental data.
Although even tiny ripples were surprisingly visible to the eye
 when in motion, 
 it was less easy to get good experimental images.

\begin{figure}
\centering
\includegraphics[angle=-0,width=0.90\columnwidth]{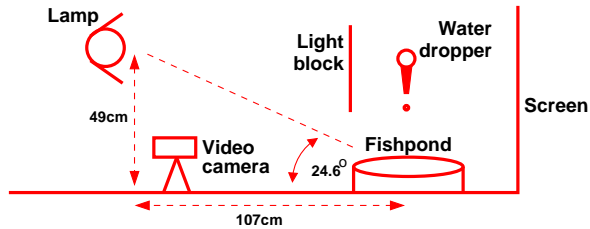}
\caption{
The experimental setup of NK and TT.
The water dropper was held in place by a retort stand
 (not shown). 
}
\label{fig-fishpond-expt-TT}
\end{figure}

\begin{figure}
\centering
\includegraphics[angle=-90,width=0.90\columnwidth]{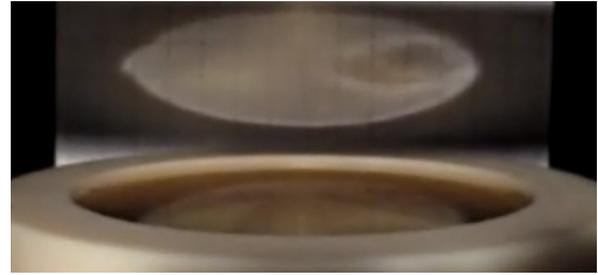}
\caption{
A snapshot of the NK/TT experiment in progress, 
 with the Fishpond in the foreground, 
 and the screen above and behind.
We can clearly see the reflected ripple patten on the right of the screen; 
 this image was taken just before the second reformation.
}
\label{fig-fishpond-expt-NKTT}
\end{figure}

Being the first attempt, 
 the NK/TT experimental setup was relatively simple.
In a darkroom, 
 they reflected lamplight off the water surface onto a screen,
 and took video images of the screen
 (see fig. \ref{fig-fishpond-expt-TT}).
This screen was shielded from any direct light from the source lamp.
The rippled water surface caused intensity variations
 on the screen, 
 depending on whether the particular perturbation 
 tended to focus or defocus the light, 
 making even very shallow ripples visible
 (see fig. \ref{fig-fishpond-expt-NKTT}).
The intensity pattern then indicated the progress of the ripples
 from source to image, 
 and back.
They then analysed the video by eye, 
 frame by frame, 
 to locate the times and positions of the reformations.

\begin{figure}
\centering
\includegraphics[angle=-0,width=0.60\columnwidth]{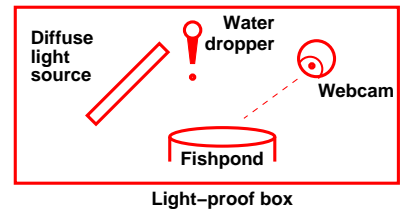}
\caption{
The experimental setup of CT and JT.
In addition the Fishpond was placed in a waterbath  
 with a heat pump,
 to enable the temperature to be changed in a controllable way.
}
\label{fig-fishpond-expt-CTJT}
\end{figure}

The CT/JT setup used lightproof box to eliminate stray light.
Inside the box they imaged the reflection of a diffuse light source
 directly using a high resolution webcam
 (see fig. \ref{fig-fishpond-expt-CTJT}).
This enabled them to electronically process the images.

In both experiments, 
 ripples were created using a water dropper
 to drop a single water droplet into the Fishpond; 
 the height of which was varied to ensure
 that the strongest possible ripples were generated, 
 but not so big as to be accompanied by splashing or bubbles.
A full range of starting positions was investigated, 
 since the Fishpond should create reformations from any point --
 although starting positions near the wall suffered due
 to edge effects.

The positioning of screens, light sources, cameras, 
 and so on were systematically varied to achieve the best images.
For NK/TT,
 a shallow angle of reflection enhanced the images, 
 although if too shallow this significantly reduced the fraction
 of the water surface that could be seen.
For CT/JT, 
 the diffuse light source was placed at a low angle, 
 but with the camera at a 90$^\circ$ reflectance angle; 
 thus giving a strong contrast from the light reflected
 off the ripples.
The diffuse light source avoided problems caused 
 by reflections off the bottom centre of the Fishpond.

\begin{figure}
\centering
\includegraphics[angle=-90,width=0.90\columnwidth]{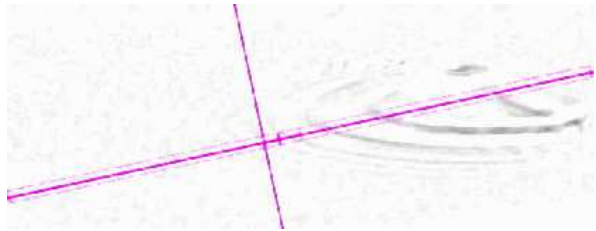} % Trant-viva-ripplescan-despeck5.ps}
\caption{
Typical differential image obtained by CT and JT, 
 after first subtracting one webcam frame from the next, 
 and them removing speckle, 
 in order to enhance ripple visibility .% based on 5x5 pixel blocks. % using xv
The purple lines cross the centre of the Fishpond, 
 with the more horizontal on a line linking 
 the source-point to the reformation point, 
 with the other at right angles from the centre of the pond.
The image is foreshortened due to the angle of view 
 of the webcam.
}
\label{fig-fishpond-CTJT-frames}
\end{figure}

%\clearpage
%-----------------------------------------------------------------------
\subsection{Setup 1}

The first experiment, 
 performed by NK and TT, 
 used the setup of fig. \ref{fig-fishpond-expt-TT},
 where videos were taken from each initial droplet
 to final dissipation of the ripples.
Each instance was viewed carefully
 frame-by-frame to determine the 
 time taken for the easily detectable 1st and 2nd reformations. 
%It should be noted that reformations occuring after the second
% were not quantitatively analysed
% because their ripples were too faint to measure
% with an acceptable error in accuracy --
% although they were still visible to the eye as a fleeting event.

Reformation times were found by first locating the desired reformation
 within a short ``target'' sequence of video frames.
With each reformation point taken to be indicated
 by the frame with the most localised ripple pattern, 
 the short target sequence was viewed independently by NK and TT, 
 both forward and backwards.
Once the best reformation frame(s) were chosen, 
 the reformation time could be calculated
 using the frame rate of the camera.
NK and TT also attempted to increase the longevity of the ripples
 in order to detect third reformations.
However, 
 reducing the water viscosity (and hence dissipation) 
 by using water at 70C 
 had negligible effect, 
 and reducing the surface tension
 using either detergent or temperature
 slowed the wave speeds, 
 increasing the effect of dissipation.

The positions of source,
 and 1st and 2nd reformation points were also compared and measured.
However, 
 since the foreshortening of the reflected ripple pattern
 means that it appears as an ellipse, 
 times were calculated from instances where the source and reformation points
 appeared on a horizontal line on the screen, 
 i.e. the long axis of the ellipse.

%-----------------------------------------------------------------------
%\subsection{Data (based on TT)}

The average times taken for first and second reformations were 
 ($1.04 \pm 0.04$)s 
 and ($1.07 \pm 0.12$)s. 
The variation between times taken for first reformations
 of different ripples at varied start positions 
 was dominated by video frame rate.
The larger variation for the second reformations
 was due to the difficulty in determining the video frame
 with the most localized ripple pattern,  
 since by then the ripples had both dispersed and diminished significantly.

The theoretically expected reformation time can be calculated
 most easily  by referring to the reference sphere.
The water depth at the centre of the Fishpond matches that 
 on the imaginary reference sphere, 
 which in our case is 2.5mm deep, 
 leading to a wave velocity of $v = \sqrt{gd} \approx 157$mm/s.
Each reformation on the sphere takes place after one half circumnavigation, 
 and the sphere radius $s$ is half that of the Fishpond radius of $r_0=100$mm; 
 thus the theoretical reformation time is 
~
\begin{align}
  T &= \frac{\pi s}{v_0}
 \approx 
  \frac{3.142 ~.~ 50\textrm{mm}}{157 \textrm{mm/s}}
 \approx 
  1.003 \textrm{s}
,
\label{eqn-expt-predictT}
\end{align}
 which seems to be in good agreement with the measurements --
 but see the discussion in section \ref{S-discuss}.

%
%-----------------------------------------------------------------------
\subsection{Setup 2}
%\subsection{Data CT/JT}

In the second experiment, 
 CT and JT directly imaged the ripples using a webcam
 and VirtualDub \cite{VirtualDub},
 as shown in fig. \ref{fig-fishpond-expt-CTJT}.
To emphasize the ripple dynamics, 
 they subtracted each frame from the previous one 
 using AviSynth \cite{AviSynth}.
The resulting differential image data
 as shown in fig. \ref{fig-fishpond-CTJT-frames}, 
 reveals only the wave motion, 
 which was then analysed using Tracker \cite{Brown-Tracker}.
Each image was scanned along the axis between source and image points, 
 giving a 2D dataset, 
 comprising a time series of 1D datasets along this axis.
The result was processed and plotted using a variety of software, 
 including Microsoft Excel, 
 Matlab \cite{MatLab},
 and Scilab \cite{Scilab}.

In order to optimise the reformation process, 
 CT and JT systematically analysed results taken for a range
 of temperatures and fill volumes, 
 as shown in tables \ref{table-volume} and \ref{table-temperature}.
The optimum temperature was found to be 15C:
 although the water viscocity (and loss) increases for lower temperatures, 
 the surface tension increases, 
 leading to faster wave speeds.
Note that the optimum fill volume centred around 190ml, 
 this can be compared to that for the design parameters, 
 which radial integration of the design depth profile is found to be 183ml.
This is in good agreement -- 
 a 3ml change in fill volume corresponds to about a 0.1mm depth change, 
 and velocity shifts of less than 2\%.
The experimental optimum filling volume of 190ml
 is higher than the design volume,
 perhaps because of the way the Fishpond fills --
 e.g.
 the design takes no account of surface tension.

% {Tdata15CeditCf-wrapper}
% {Tdata20CeditCf-wrapper}

\begin{figure}
\centering
\includegraphics[angle=-0,width=0.84\columnwidth]{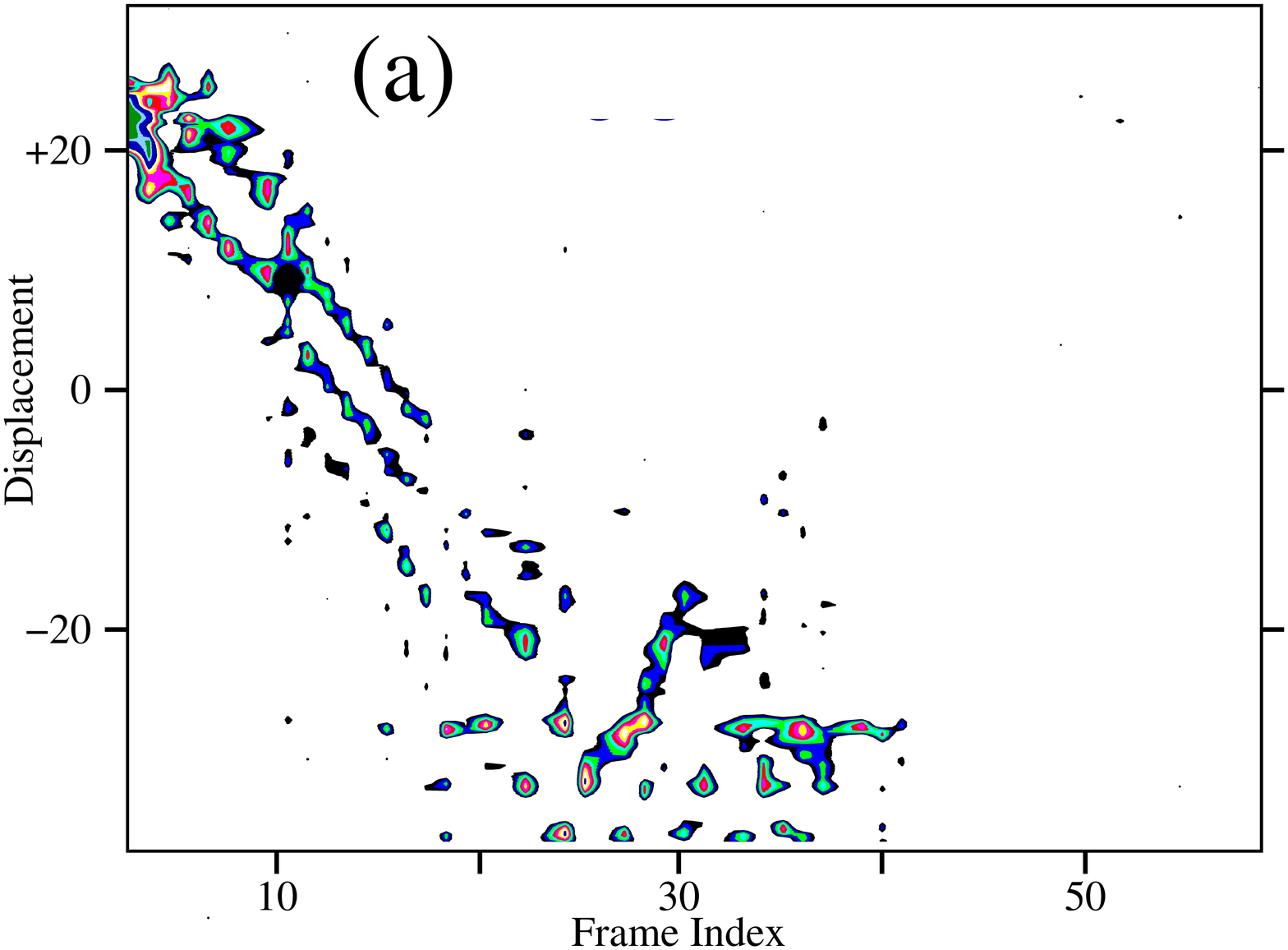}\\
~\\
\includegraphics[angle=-0,width=0.84\columnwidth]{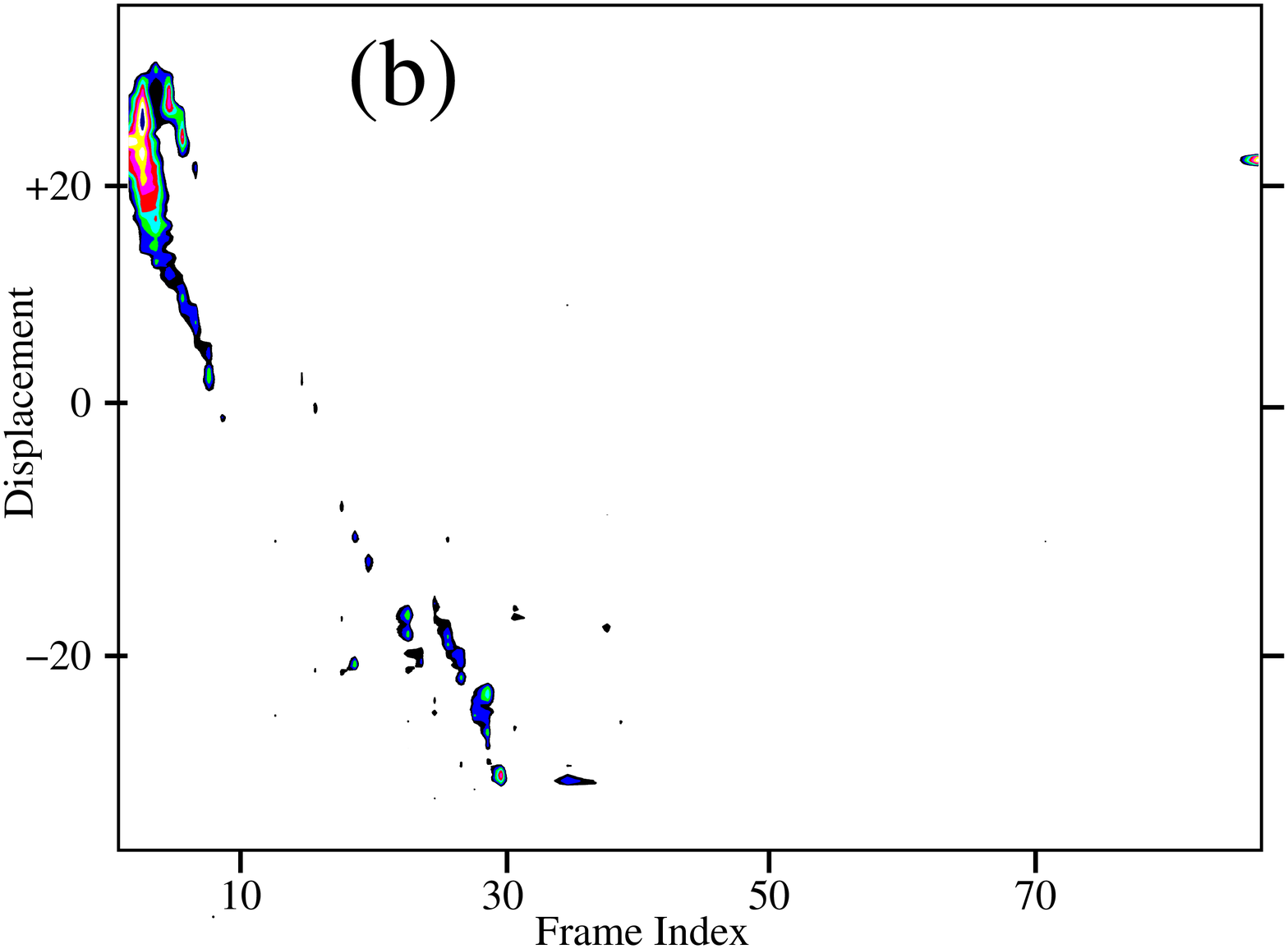}
\caption{
Differential luminance data 
 indicating the presence of ripples along 
 the axis between initial disturbance and reformation point.
As time (and frame index) progresses, 
 we see the ripples travel from source to reformation, 
 although the pattern is complicated by the two possible paths -- 
 either with an early reflection off the bowl edge, 
 or with a late reflection.
This is in addition to the spreading out of the ripple pattern
 due to dispersion.
%This data was taken for water at 15C.
Results for two different water temperatures are shown, 
 at both (a) 15C, and (b) 20C, 
 with 15C water giving better data, 
 in agreement with Table \ref{table-temperature}.
These contour plots are made by averaging over 
 adjacent points 
 and using a logarithmic scale; 
 the contours are evenly spaced, 
 with a minimum level chosen to best display the ripple patterns.
}
\label{fig-fishpond-CTJT-lum}
\end{figure}

%
%Scaling 
%    +"20" to "-20" is 106mm in xfig
%    top to bottom  is 188mm in xfig 
% hence 188mm is really 200mm, ie fishpond diameter
% hence "20" is really 20 * 200/188 --> 21.27mm
% hence double-20 spacing should be at 106 * 188/200 = 99.6 
% and +20, -20 at +10, -10

In fig. \ref{fig-fishpond-CTJT-lum}(a) we see the first reformation 
 very clearly, 
 although dispersion has spread out the initial impulse
 introduce by a falling water drop.
On the first traversal, 
 we can not only see the evidence of several ripple crests, 
 but also the slight fanning as the ripples disperse.
Further reformations, 
 although apparent to the eye, 
 do not show up over the imaging noise.

\begin{table}[h]
 \begin{tabular}{|l|c|c|c| c|c|c| c|}
  \hline
    Volume [ml]  &  170 &  175 &  180 &  185 &  190 &  195 &  200 \\
  \hline
    Reformations &    2 &    2 &    2 &    3 &    3 &    3 &    2 \\
  \hline
 \end{tabular}
\caption{Relationship between water volume
 in the Fishpond 
 and the number of reformations visible to the eye.}
\label{table-volume}
\end{table}

\begin{table}[h]
 \begin{tabular}{|l|c|c|c| c|c|c| c|c|c|}
  \hline
    Temperature [C] &  0 &  5 & 10 & 15 & 20 & 25 & 30 & 35 & 40 \\
  \hline
    Reformations    &  1 &  2 &  2 &  3 &  2 &  2 &  2 &  2 &  2 \\
  \hline
 \end{tabular}
\caption{Relationship between temperature 
 of the Fishpond
 and the number of reformations visible to the eye.}
\label{table-temperature}
\end{table}

Reformation times and errors were extracted from data like that shown on 
 fig. \ref{fig-fishpond-CTJT-lum}
 using a curve fitting process.
For example, 
 at the optimum temperature of 15C, 
 the first reformation was calculated to have occured
 at the 23rd frame (at $0.72 \pm 0.06$s), 
 with the second being 30 frames later (+ $1.03 \pm 0.11$s).
The second reformation was notably slower than the first, 
 but then the error is also much larger; 
 but of course some slowing might be expected since the longer wavelengths
 both persist longer and travel more slowly.

%As noted already, 
% the reformation time including surface tension effects is faster 
% than that of about 1s 
% calculated in eqn. \eqref{eqn-expt-predictT} by some 30\%; 
% in quite good agreement with the measured value.

Finally, 
 despite the difficulty in extracting second and third reformation times
 from the video data, 
 and in seeing them in plots such as fig. \ref{fig-fishpond-CTJT-lum}, 
 in the original videos themselves the third reformation is clearly visible
 to the eye.
This suggests that significant performance improvements 
 are still possible in the automated processing of the data.

%
% =======================================================================
\section{Surface tension}\label{S-discuss}

Since water is the obvious liquid to use in the Fishpond, 
 and it has a significant surface tension, 
 we should estimate its effects.
The wave velocity including the effects of surface tension $\sigma$
 (in N/m) on waves of wavelength $\lambda = 2\pi/k$, 
 in a fluid of density $\rho$ and depth $d$
 is
~
\begin{align}
  v^2
 =
  \frac{\omega^2}{k^2}
&=
  \frac{g}{k}
  \left[
    1
   +
    \frac{\sigma k^2}{g \rho} 
  \right]
  \tanh
    \left(
      k d
    \right)
.
\end{align}
Thus the correction to the leading term 
 which gives us the shallow water wave speed
 is a factor $\epsilon = \sigma k^2/g\rho$.
Surface tension in water reduces with temperature, 
 being about 0.073 N/m at 20C, 
 but 0.061 N/m at 90C.
At about 20C,
~
\begin{align}
  \epsilon
=
  \frac{\sigma k^2}
       {g \rho} 
&\approx
  \frac{0.073 k^2}
       {9.81 \times 1000}
=
 7.44 \times 10^{-6} k^2
. 
\end{align}
For water waves of wavelength 20mm, 
 $k = 2\pi/0.02 \approx 314$m$^{-1}$, 
 so that 
 $\epsilon \approx 0.73$; 
 the wave speed is therefore 
 a factor of $\sqrt{1.73}$ (or 30\%) higher
 than expected based on depth alone, 
 and is wavelength dependent even in the shallow water limit.
However, 
 this speed shift is not \emph{depth} dependent, 
 so the refocussing character of the Fishpond is unaffacted --
 but different wavelengths reform at different times.
This dispersion means that a determination of the reformation time
 becomes harder as time progesses, 
 and will depend on the specific details of how
 a given reformation time is evaluated.

Initially, 
 the NK/TT measured reformation time of $(1.04 \pm 0.04)$s
 seemed in good agreement
 with the simple prediction of eqn. \eqref{eqn-expt-predictT}, 
 i.e. 1.00s.
However, 
 we can now see that 
 surface tension effects should reduce this prediction by 30\%
 to about 0.7s, 
 which agrees with that measured by CT/JT, 
 and not that of NK/TT.

But why do the two experiments
 give such different outcomes?
Two scenarios, 
 which are not mutually exclusive, 
 suggest themselves.

First, 
 the criteria for choosing the reformation times differed, 
 and this will affect which frame of video selected --
 NK/TT chose by eye the frame with the smallest region of disturbed water, 
 whereas CT/JT applied a simple fitting algorithm to digitised data
 along one axis.

Second, 
 NK/TT relied upon the reported frame rate of their camera, 
 and perhaps this was not reliable; 
 although with hindsight we realise that 
 their framerates might have been easily calibrated by videoing a clock
 either before or at the same time as the each experimental run.

It is gratifying that the more sophisticated setup of CT/JT 
 gives good agreement with theory, 
 although it is not clear why the first attempt by NK/TT 
 did less well.
Neverthless, 
 one of the features of this student project
 was it could be implemented in many different ways -- 
 the students were given the Fishpond, 
 some reading material, 
 and some suggestions and then 
 largely left to get on with it as independently they wished.
Still other experimental set-ups and measurements are possible, 
 and so we expect that the Fishpond itself
 will be reused many times in the future.

%http://www.engineeringtoolbox.com/water-surface-tension-d_597.html

%
% =======================================================================
\section{Summary}\label{S-summary}

We have shown how an exotic phenomenon from transformation optics --
 the Maxwell's fisheye lens --
 can be converted into simple water waves
 in a tabletop ``Maxwell's Fishpond''.
This is currently being used sucessfully as a third year undergraduate
 experimental project in the Physics Department of Imperial College London.
While the remarkable series of image reformations
 provides the hook which makes the project interesting, 
 there are many other features that can be investigated as part
 of the experiment.
Most straightforwardly, 
 there is a variety of imaging possibilities to be investigated 
 (two of which were discussed here), 
 and various experimental conditions -- 
 lighting, fill depth, etc --
 to be determined.
Also, 
 the effect of viscosity on performance
 can be tested by changing the water temperature, 
 or surface tension can be removed by adding detergent --
 or other liquids might be used.
A transparent Fishpond might be made
 so as to image the ripples in transmission, 
 or a vibrating source could be used in an attempt
 to generate standing waves.
For the more mathematically inclined, 
 the nature of stereographic projections can be researched, 
 other comparable devices -- 
 e.g. the Eaton or Luneburg lenses --
 considered, 
 or numerical simulations attempted.
Alternatively, 
 rather than only aiming to optimise the number of reformations,
 or visibility to the eye, 
 but it is also possible to consider ease or simplicity of fabrication, 
 with a view to testing performance as a function of size.
Since our simulations show that the general behaviour persists
 even for an approximate depth profile -- 
 such as a shallow dome --
 sophisticated or precise manufacturing processes are not needed.

In this way, 
 this simple, 
 eye-catching device provides a rich playground
 in which a wide variety of students can test their skills
 while investigating a novel device not only part of contemporary research --
 that of transformation optics and acoustics -- 
 but with a history that goes back to Maxwell himself.

%
% =======================================================================
\section*{Appendix}\label{S-appendix}

%
% -----------------------------------------------------------------------
\subsection{Other Lenses}\label{S-fish-EatonEtc}

One might also construct Eaton and Luneburg lenses\cite{Luneberg-MTO},
 or even their generalizations \cite{Tyc-HSB-2011njp,Sarbort-T-2012jo},
 using water waves.
The refractive index profiles, 
 which are proportional to 
 the inverse of the velocity profiles
 for the Eaton and Luneburg lenses, 
 are
 $n_{\textrm{Eaton}}(r) = \sqrt{(2r_0-r)/r}$
 and 
 $n_{\textrm{Luneburg}}(r) = \sqrt{(2r_0^2-r^2)/r_0^2}$. 
Then, 
 by comparing velocity profiles between the optical and water wave cases, 
 and choosing a reference depth $d_0$ and reference radius $r_0$, 
 we find that the \emph{retro-reflecting} Eaton pond feature
 and the 
 \emph{focussing} Luneburg pond feature
 need the depth profiles
~
\begin{eqnarray}
  d_{\textrm{Eaton}}(r) 
&=
  \frac{r d_0}{2 r_0 - r}
,\\
%\end{eqnarray}
% whereas the \emph{focussing} Luneburg pond feature
% needs a depth profile with
%~
%\begin{eqnarray}
  d_{\textrm{Luneburg}}(r) 
&=
  \frac{r_0^2 d_0}{2 r_0^2 - r^2}
.
\end{eqnarray}

%
% -----------------------------------------------------------------------
\subsection{Other waves}\label{S-fish-other}

It is possible to imagine other implementations of the Maxwell fisheye
 concept.
For example, 
 an adaption of the expertise demonstrated 
 by Bramhavar et al in \cite{Bramhavar-PNENM-2011prb}
 might give rise to an appropriately tapered ``Maxwell's Platter''
 with the same behaviour for acoustic waves
 in a solid.

%
% -----------------------------------------------------------------------
\subsection{Finite Element simulations}\label{S-model-OpenFOAM}

\begin{figure}
\includegraphics[angle=-0,width=0.95\columnwidth]{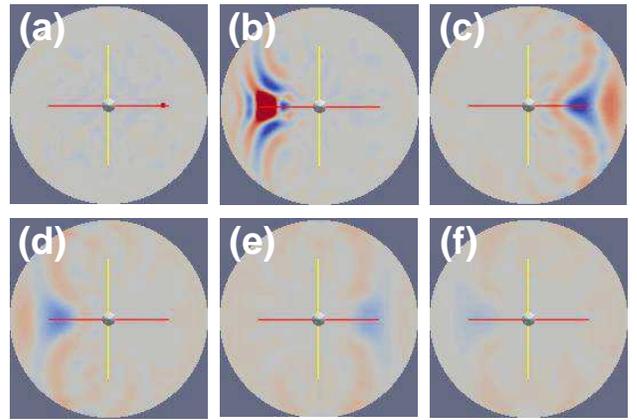}
\caption{
Snapshots from a Maxwell's Fishpond 200mm diameter and 10mm deep,
 simulated using OpenFOAM,
 at the start and each subsequent reformation
 up to the fifth.
%The numerical mesh has imprinted on-diagonal features 
% visible at 0.0875s, 
% also the concentric ripples were slightly distorted, 
% impacting on the precision of the reformations.
Although the simulation retains to a reasonable extent the 
 repeated refocussing, 
 the wave dispersion continually increases the length of the pulse of ripples, 
 so that reformations, 
 while still remaining on the scale of a wavelength, 
 have an ever increasing duration.
}
\label{fig-foampond-snapshots}
\end{figure}

As an estimator of the necesarily imperfect fishpond experiment, 
 we (PK) also did more realistic finite element simulations
 using the open source simulator OpenFOAM \cite{OpenFOAM}
 using the {interMixingFoam} engine, 
 on a fast desktop PC.
%These simulations used a customised block meshing for the fishpond, 
% because we wished to specify the depth profile at high resolution.
Despite computational constraints, 
% the mesh could not be made as fine as we would like,
 the general character of the idealised process was preserved, 
 and the effect of dispersion demonstrated.
A typical simulation result is shown on fig. \ref{fig-foampond-snapshots}; 
 others indicate that larger fishponds may perform better than ours -- 
 although will be harder to construct, 
 and will suffer more from dispersion.

%
% -----------------------------------------------------------------------
\subsection{MEEP ctl files}\label{S-meepctl}

\begin{widetext}

\begin{verbatim}

; Simulate the ideal Maxwell Fishpond and approximate circular Ponds 
; by converting a depth profile into a refractive index.
;
; 
; Dr Paul Kinsler, 2011 & 2012
; 
; 1) an exact & idealised Maxwell's Fishpond
; 2) an approximate "SC" Maxwell's Fishpond whose water-depth profile is  
;    determined by a shallow, convex spherical cap, with a min:max 
;    depth ratio of 1:4 to match that of an exact Fishpond.
; 3) a Fishpond with a constant depth
; 
; The shallow water-wave speed resulting from the depth profile is converted 
; into a refractive index.
; 
; ======================================================================
; Fisheye sizes and parameters
; 
(define-param n  2) ; base index of fisheye
(define-param w  1) ; width of waveguide
(define-param r 10) ; inner radius of ring

; ======================================================================
; Computational resolutions etc
;
(define-param pad 2) ; padding between waveguide and edge of PML
(define-param dpml 2) ; thickness of PML

(define sxy (+ 0.1 (* 2 (+ r w pad dpml)))) ; cell size !odd!
(set! geometry-lattice (make lattice (size sxy sxy no-size)))

; ======================================================================
; Refractive index profile functions 
;  to calculate the local refractive index as it varies with position.
;
;  rr = sqr(x^2+y^2)
(define (frr p)
        (sqrt (+ (* (vector3-y p) (vector3-y p)) 
                 (* (vector3-x p) (vector3-x p))
              )
        )
)

; --------------------------------------------------------------------
; 1) The Maxwell's Fishpond, 
;            and the variation with position of its local refractive index
; 
;  1 + rr^2/r^2
(define (fdivisor rr)
   (+   1   ( / (* rr rr) (* r r) )
)  )

(define (refindex rr)
   (/ n (fdivisor rr))
)


; --------------------------------------------------------------------
; 2) The approximate spherical-cap "SC" Fishpond, 
;            and the variation with position of its local refractive index
; 
; calculate the sphere-size for an r-radius pond with 1:4 depth ratio
; R=10 => D = R^2/6 + 3/2 = 18 1/6 = 18.166666667

(define-param DD   (+ 1.5 (/ (* r r) 6)))
(define-param DDp1 (+ 1 DD))
(define-param DD2  (* DD DD))

;  1+D - sqrt(D^2-rr^2)
(define (fdivisorSC rr)
   (- DDp1 (sqrt (- DD2 (* rr rr) )))
)

(define (refindexSC rr)
   (/ n (sqrt (fdivisorSC rr)))
)

; --------------------------------------------------------------------
; 3) The constant-depth Fishpond
;            and the non-variation with position of its local refractive index
; 
(define (refindexCO rr)
   (sqrt (sqrt 2))
)


; --------------------------------------------------------------------
; --------------------------------------------------------------------
; convert refractive index to epsilon, and make the correct medium
; by uncommenting ONE of the allowed refractive index profile 
; function calls in (define (eps rr) ...)


(define (eps rr)
   (* (refindex rr) (refindex rr))    ; uncomment for exact Fishpond
 ;  (* (refindex rr) (refindexSC rr))  ; uncomment for approx SC Fishpond
 ;  (* (refindex rr) (refindexCO rr))  ; uncomment for constant-depth Fishpond
)

; Definition of the medium f(p)
(define (fmedium p)
        (make medium
                (epsilon (eps (frr p)))
)       )




; ======================================================================
; 
; Create the pseudo-Fisheye/Fishpond structure
;


; Create a ring waveguide by two overlapping cylinders - later objects
; take precedence over earlier objects, so we put the outer cylinder first.
; and the inner (air) cylinder second.
(set! geometry 
  (list 
        (make cylinder (center 0 0) (height infinity)  
                       (radius (+ r w)) (material metal))
        (make cylinder (center 0 0) (height infinity) 
                       (radius r) 
                       (material (make material-function 
                                       (material-func fmedium) )))
) )



; ======================================================================
; Set up the PML at the simulation boundaries
;

(set! pml-layers (list (make pml (thickness dpml))))
(set-param! resolution 10)

; ======================================================================
; 
; SOURCES: Put a single point source on the y-axis at x=7.20um
; 
; 

(set! sources 
  (list
    (make source
      (src (make gaussian-src (frequency 0.333333)  (width 3))) ; was 33
      (component Ez) (center 5.00 0.00) ; was 7.2
    )
  )
)

; ======================================================================
; 
; Run the simulation
; 
; 
(run-until 180
           (at-beginning output-epsilon)
           (to-appended "I" 
             (at-every 0.125 (synchronized-magnetic output-tot-pwr))
           )
           (to-appended "ez" 
             (at-every 0.125 output-efield-z))
           )

\end{verbatim}
\end{widetext}

%
% =======================================================================
%\bibliography{/home/physics/_work/bibtex.bib}

%
% =======================================================================
\end{document}